# Optically Coherent Nitrogen-Vacancy Centers in HPHT Treated Diamonds

**Authors:** Yuan-Han Tang[1,2,#], Xiaoran Zhang[3,4,5,#], Kang-Yuan Liu[1,2], Fan Xia[1,2], Huijie Zheng[1], Xiaobing Liu[3,4,*], Xin-Yu Pan[1,6,7], Heng Fan[1,2,6], Gang-Qin Liu[1,6,7,*]

**Affiliations:**

[1] Beijing National Laboratory for Condensed Matter Physics, and Institute of Physics, Chinese Academy of Sciences, Beijing 100190, China.

[2] School of Physical Science, University of Chinese Academy of Sciences, Beijing 100049, China.

[3] Laboratory of High Pressure Physics and Material Science, School of Physics and Physical Engineering, Qufu Normal University, Qufu, Shandong Province, 273165, China.

[4] Advanced Research Institute of Multidisciplinary Sciences, Qufu Normal University, Qufu, Shandong Province, 273165, China.

[5] School of Mathematical Sciences, Qufu Normal University, Qufu, Shandong Province, 273165, China.

[6] CAS Center of Excellence in Topological Quantum Computation, Beijing 100190, China

[7] Songshan Lake Materials Laboratory, Dongguan, Guangdong 523808, China.

[#] These authors contributed equally

[*] Correspondence should be addressed to X.B.L. (xiaobing.phy@qfnu.edu.cn) or G.Q.L. (gqliu@iphy.ac.cn)

**Abstract:** As a point defect with unique spin and optical properties, nitrogen-vacancy (NV) center in diamond has attracted much attention in the fields of quantum sensing, quantum simulation, and quantum networks. The optical properties of an NV center are crucial for all these quantum applications. However, NV centers fabricated by destructive methods such as electron irradiation or ion implantation usually exhibit poor optical coherence. In this work, we demonstrate a non-destructive method to fabricate optically coherent NV centers. High-purity single crystal diamonds are annealed under high pressure and high temperature (>1700 °C, 5.5 GPa), and individually resolvable NV centers with narrow PLE linewidth (<100 MHz) are produced. The high-pressure condition prevents the conversion of diamond to graphite during high-temperature annealing, significantly expanding the parameter space for creating high-performance artificial defects for quantum information science. These findings deepen our understanding of NV center formation in diamond and have implications for the optimization of color centers in solids, including silicon carbide and hexagonal boron nitride.

**Main Text:**

Optical coherence is at the heart of all quantum optics and quantum information processing applications. For nitrogen-vacancy (NV) centers in diamond, the optical properties are determined by the lifetime of the excited state, as well as by the fluctuation of their local environments, including the local lattice strain and electronic field [1]. The latter have attracted much attention in recent years. On the one hand, low strain and stable charge environments are prerequisites to realize high-fidelity single-shot readout of the NV spin states [2], to establish entanglement between distant NV centers [3–5], and to boost the sensitivity of NV-based quantum sensors [6]. On the other hand, as a point defect in the diamond lattice, the presence of an NV center is usually associated with other defects, e.g. substitutional nitrogen atoms (P1 centers) and other electron donors, which could help stabilize the charge state of an NV center ($NV^-$ is preferred) [7]. Under laser excitation, the ionization and electron capture process of an NV center occurs randomly, leading to unavoidable fluctuation of its local charge environment and resulting in the spectra diffusion of NV centers [8–11]. Ideally, there should be no or only one electron donor in the vicinity, so that the charge environment of the NV center remains the same under laser excitation. However, such a perfect lattice is practically rare to find. An alternative approach is to reduce the local defects around the NV center.

There are several strategies for producing optically coherent NV centers in diamond. The simplest method is to use NV centers that are naturally formed during chemical vapor deposition (CVD) of high-purity diamond, which have the best optical properties to date. The spectral diffusion width of these NV centers is less than 100 MHz, enabling several milestone experiments with this system [2,3,5,12,13]. Moreover, NV centers fabricated by laser writing also show good optical coherence and have the advantage of sub micrometer precision positioning [14]. Alternatively, recent works have shown that electron irradiation and subsequent high-temperature annealing [15] of high-purity diamond, or simply high-temperature annealing can produce optically coherent NV centers [16]. Meanwhile, whether ion implantation can form optically stable NV centers is still under debate [9,10,17,18]. Overall, although natural NV centers are perfect, they are rare and therefore do not meet the requirements of quantum applications, which are growing day by day. NV centers produced by destructive methods such as laser writing and electron irradiation are always compromised by accompanying defects (mostly vacancies).

In this work, guided by the idea of forming NV centers with minimal local defects in the vicinity, we demonstrate a simple, massive and non-destructive method for the fabrication of optically coherent diamond NV centers. Two high purity single-crystal diamonds are annealed (without electron irradiation or ion implantation) at temperatures of 1750 °C and 1800 °C, with a high-pressure protection of 5.5 GPa to effectively prevent graphitization of the diamonds. Dispersed NV centers with excellent optical coherence are formed in both samples. We systematically study the optical and spin properties of these NV centers, which are suitable for the most stringent quantum applications.

**Experimental results**

High-purity <100> oriented single-crystal diamonds grown by chemical vapor deposition (CVD) are used (nitrogen concentration below 5 ppb, Element Six). HPHT annealing is carried out in a China-type cubic high-pressure apparatus (SPD-6×1200), at temperatures of 1750 °C (sample S1) and 1800 °C (sample S2) under a pressure of 5.5 GPa. The process of HPHT annealing treatment is shown in Fig. 1(a). The temperature is ramped to the target temperature within 5 minutes after the pressure is increased to 5.5 GPa. The temperature of the sample is maintained for 2 hours and then cooled to room temperature within 10 minutes. No graphitization takes place during this process. Our previous work has shown that at a pressure of about 5 GPa, high temperature annealing of diamond at up to 2000 °C can be performed safely and without graphitization [19]. HPHT annealing has long been used to improve the photoluminescence (PL) and spin properties of NV centers in diamond [20], but the optical coherence of NV centers in HPHT treated diamonds has not been fully investigated [18,21].

The confocal images of the sample are taken before and after HPHT treatment to track the formation of NV centers, as shown in Fig. 1(b-d). Before HPHT annealing, no NV centers are detected in these high-purity diamonds. When the annealing temperature reaches 600 °C, the vacancies in the diamond lattice start to move [22–24], and the N atoms start to move above 1500 °C [16]. Although the speed of defect migration in the lattice is very low, e.g., the diffusion constant of vacancies at 1000 °C is only $1.5 \times 10^{-14}$ cm$^2$s$^{-1}$ [25], there is a probability that NV centers will form if there is enough time for the defects to diffuse. In our experiment, no NV center appeared after the first two-hour annealing at 1700 ℃, so we heated up to 1750 °C (sample S1) and 1800 °C (sample S2) for a further two hours. After HPHT annealing, a large number of NV centers appeared in both diamond samples, as shown in Fig. 1(c) and (d). Higher annealing temperature can produce more NV centers. The density is estimated to be ~0.04 /μm$^3$ and ~0.4 /μm$^3$ for sample S1 (1750 °C, 2 hours) and sample S2 (1800 °C, 2 hours), respectively. We examine different areas of the diamond sample and at different depths, from 5 μm to 230 μm below the diamond surface. At all measured depths, dispersed NV centers with similar density are observed (see confocal images in Fig. S3 of the Supplemental Material), suggesting that the formation of NV centers is due to the combination of intrinsic vacancies and nitrogen atoms in the diamond lattice, rather than vacancies from the diamond surface during HPHT treatment.

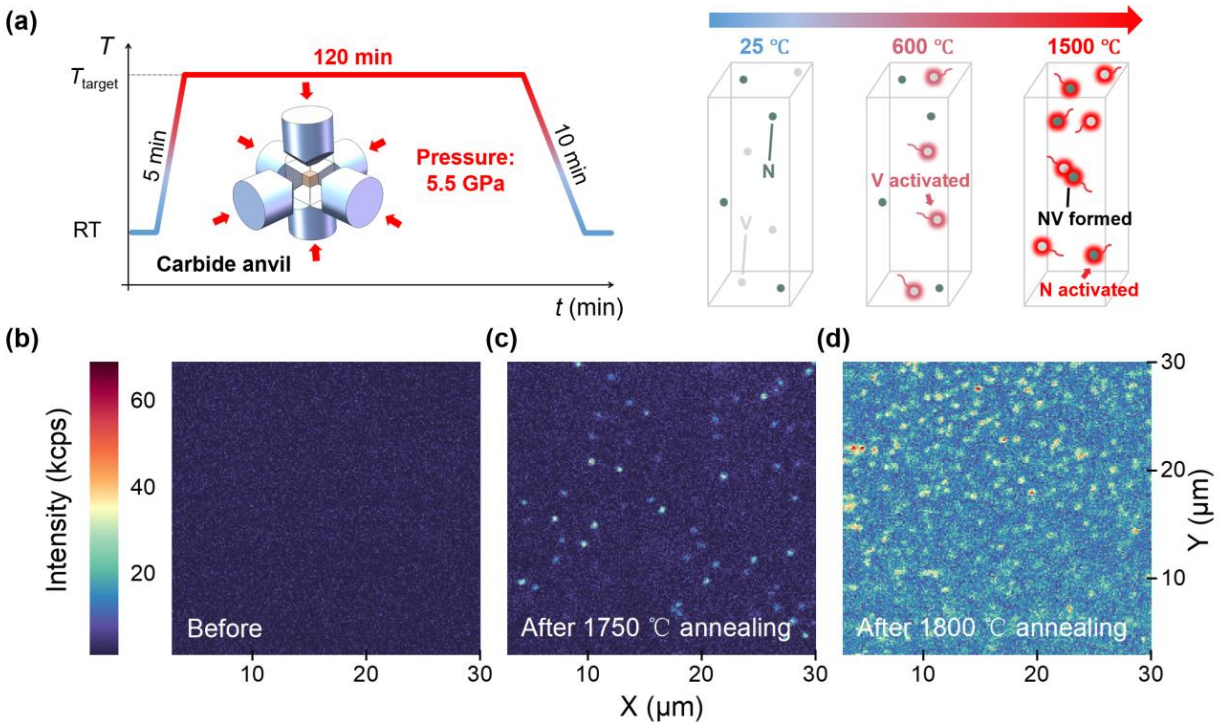

**FIG. 1. Formation of optically coherent NV centers by HPHT annealing.** (a) The HPHT annealing process. When the annealing temperature reaches 600 °C, the vacancies in the diamond lattice start to move, and the N atoms start to move above 1500 °C. The two combine to form the NV center. (b-d) 2D confocal images of the diamond under study before (b) and after HPHT treatment at (c) 1750 °C and (d) 1800 °C. Before HPHT annealing, no NV center is observed in the high-purity diamond (N concentration < 5 ppb, E6). Single NV centers are formed after HPHT treatment of both samples, and denser NV centers are found in the diamond treated at 1800 °C @ 5.5 GPa (sample S2).

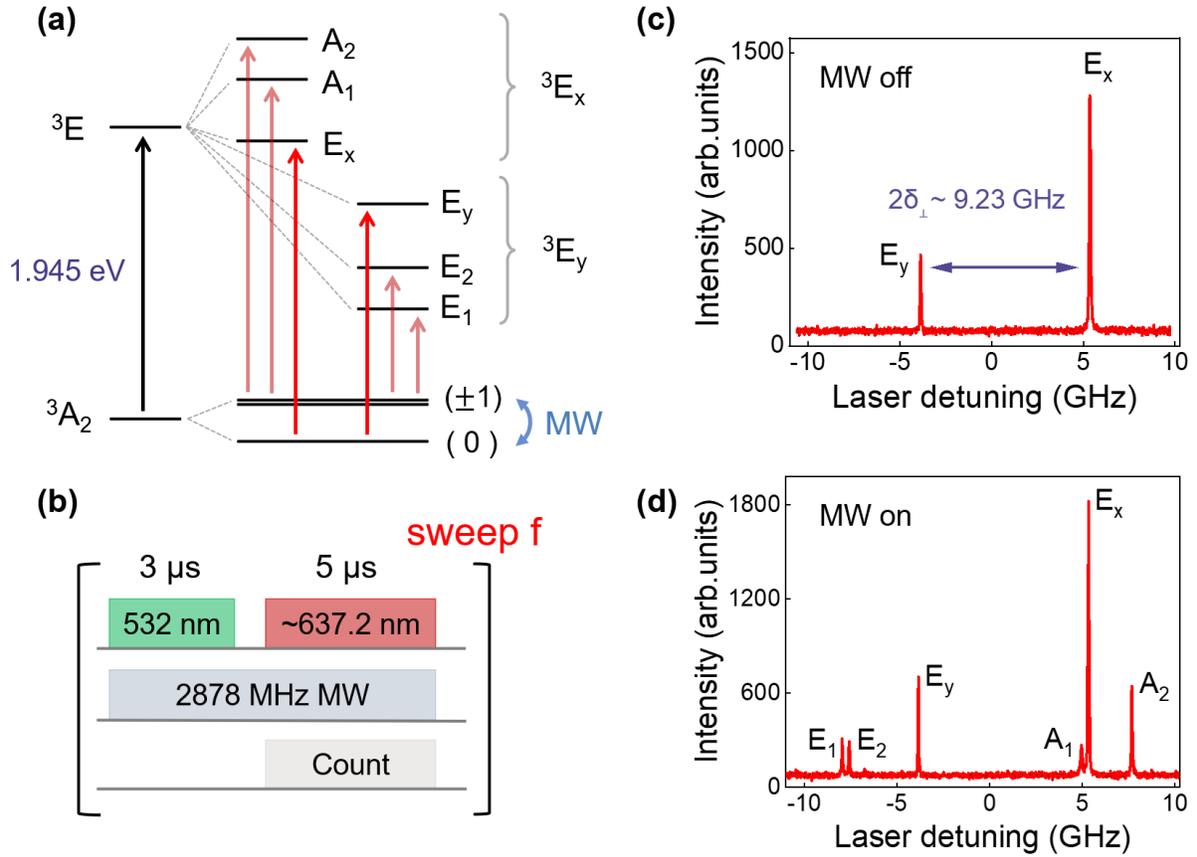

**FIG. 2. PLE spectra of NV centers generated by HPHT annealing.** (a) Energy level structure of an NV center. At low temperatures, the excited state of NV is an orbital doublet and spin triplet. (b) Experimental sequence for measuring the PLE spectra. The first green laser pulse sets the NV center to the minus charge state, then a near-resonant laser pulse (~637.2 nm) is applied and fluorescence at 650 nm-800 nm is collected. To flip the spin state of the NV center, a resonant microwave is applied (2.878 GHz, optional). (c-d) Typical PLE spectra of the $m_s=0$ state (c) and $m_s=0, \pm 1$ states (d) of an NV center in the sample S1. The split between $E_x$ and $E_y$ transitions is 9.23 GHz and the linewidths are $56 \pm 2$ MHz and $74 \pm 6$ MHz for the $E_y$ and $E_x$ transitions, respectively.

The optical coherence of the generated NV centers is characterized by the photoluminescence excitation (PLE) measurement. Figure 2 (a) shows the energy level diagram of an NV center. At low temperatures, the excited states, which consist of a spin triplet and an orbital doublet and are referred to as $^3E_x$ and $^3E_y$, can be well resolved [26,27]. The position of the six energy levels is determined by the spin-orbit interaction, the spin-spin interaction, and the local transverse strain. The PLE spectral data are recorded with a tunable red laser sweeping across the ZPL peak of the NV center to-be-measured (~637.2 nm) at 4.8 K. The experimental pulse sequence is shown in Fig. 2 (b). The power of the red laser is set to the saturation power, 620 nW, of this NV center (see Fig. S2 in Supplemental Material). The fluorescence signal of the phonon sideband (PSB, 650 ~ 800 nm) of the NV center is collected. Since sustained resonant excitation can lead to electron spin flip and even ionization of the NV center [2], we apply a 3-μs-532-nm laser pulse (~560 μW) before each frequency turning to set the NV center to the negative charge state.

Figure 2 (c) shows a typical PLE signal of the electron spin state $m_s$=0. The spectrum shows in Lorentzian line shapes with FWHM of 56 ± 2 MHz ($E_y$) and 74 ± 6 MHz ($E_x$). The splitting between $E_y$ and $E_x$ transitions is 9.23 GHz, which is twice the transverse strain [26]. Figure 2 (d) shows the PLE spectrum of the same NV center when 2878-MHz microwave is applied, it shows six distinct peaks corresponding to the six transitions in Fig. 2 (a). It is worth noting that despite the inevitable charge conversion by the 532-nm laser, the PLE spectrum still shows a narrow linewidth of less than 100 MHz, indicating very few defects near the NV centers. This value is comparable to the best record in the literature [9]. With such a small linewidth, the NV centers generated by HPHT annealing can be used in the most advanced quantum experiments , e.g., quantum interference of distant NV centers [3].

More NV centers are measured, and the statistical results are summarized in Fig. 3. Thirty-five NV centers are measured in sample S1, among which twenty-four show significant and stable PLE signals, two show no PLE signals, and nine show multiple peaks, suggesting that these NVs may be clustered or the PLE signals are unstable (see Fig. S3 and Fig. S4 in Supplemental Material for extended data). The linewidths of 24 NV centers in sample S1 are summarized in Fig. 3 (a). The mean value of the linewidth is 103.3 MHz, with a standard deviation (SD) of 64.2 MHz and a standard error (SE) of 13.1 MHz. As summarized in Fig. 3 (c), the empirical cumulative distribution function (ECDF) shows that the probability of linewidth less than 100 MHz is 0.542, indicating that the HPHT annealing method can efficiently generate NV with narrow linewidth. The smallest linewidth observed in the experiment is 30 ± 2 MHz, as shown in the inset of Fig. 3 (e). Twenty-four points are investigated in sample S2, twenty-two of which show multiple peaks in the PLE spectrum, and only two have two sharp peaks (similar to the spectrum in Fig. 2 (b)). This is plausible because the NV density of sample S2 is much higher than that of sample S1, leading to a high probability of finding overlapping NV centers within the laser spot.

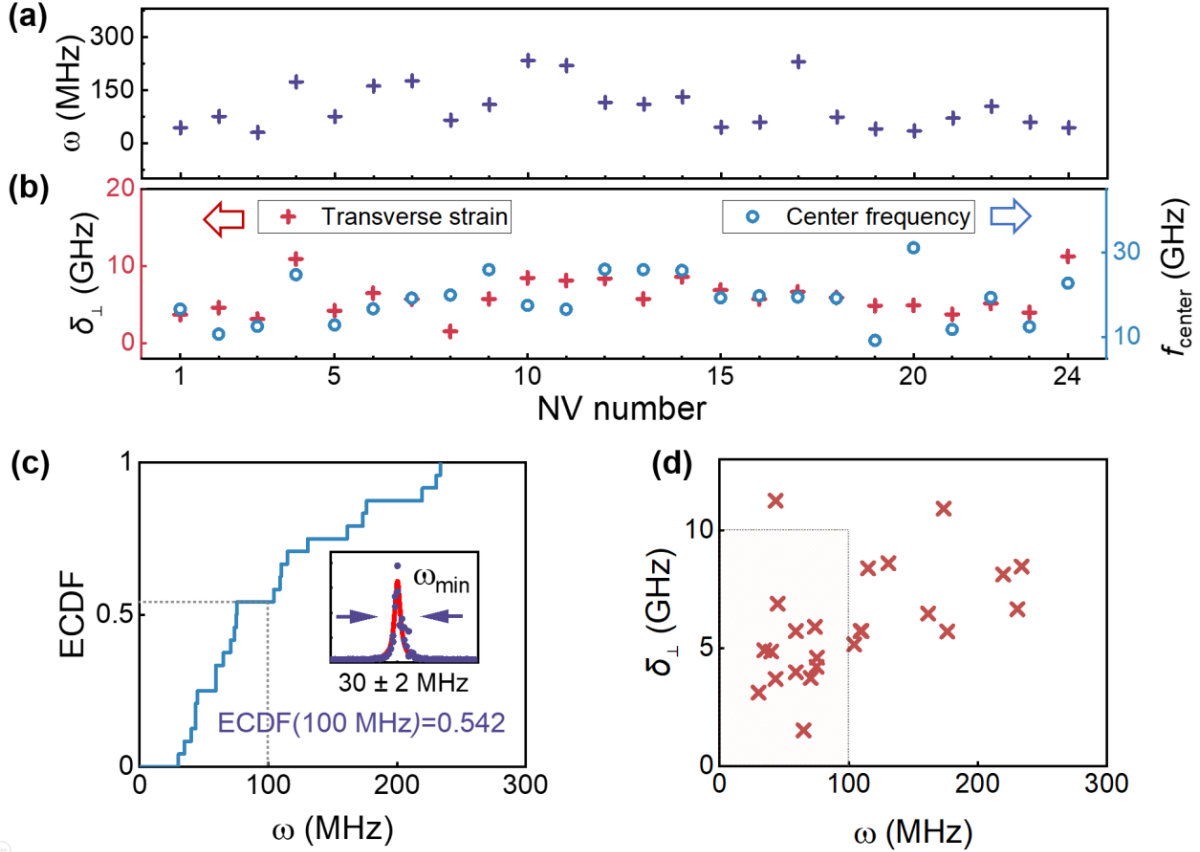

**FIG. 3. Statistical results of the NV optical properties.** (a-b) Scatter plot of (a) the linewidth $\omega$, (b) the transverse strain $\sigma_\perp$, and the center frequency $f_{\text{center}}$ (relative to 470.47 THz) of twenty-four single NV centers. These NV centers are randomly selected over the entire diamond sample (~ 2 mm). (c) The empirical cumulative distribution function (ECDF) of the linewidths in the 1750°C treated diamond (sample S1). The value of the ECDF for 100 MHz is 0.542. (d) Scatter plot of transverse strain and linewidth. The points in the shaded area represent NV centers with both low transverse strain (<10 GHz) and narrow linewidth (<100 MHz).

In addition to the PLE linewidth, we also examined the strain environment of these NV centers. Strain can be divided into axial strain along the NV axis and transverse strain perpendicular to the NV axis. The axial strain shifts all sublevels in the excited state simultaneously and can be estimated from the movement of the central frequencies of $E_x$ and $E_y$. Transverse strain splits $E_x$ and $E_y$ and causes mixing of the sublevels, the value of which is equal to half the distance between $E_x$ and $E_y$ [27]. One might be more interested in NV centers with transverse strains below 10 GHz, which exhibit low level mixing of the excited states and a long lifetime of the cyclic transitions, thus enabling high-fidelity readout of NV spin states [2] and other advanced quantum applications [3,4,13]. All twenty-four NV centers with good PLE signals in sample S1 have transverse strains of less than 12 GHz, and their center frequencies are distributed within 25 GHz (Fig. 3 (b)), indicating a uniform strain distribution throughout the sample. Of the twenty-four points, twelve have transverse strains of less than 10 GHz and linewidths of less than 100 MHz (see Fig. 3 (d)). In other words, about one-third of the thirty-five points we studied have optical coherence good enough for advanced quantum applications.

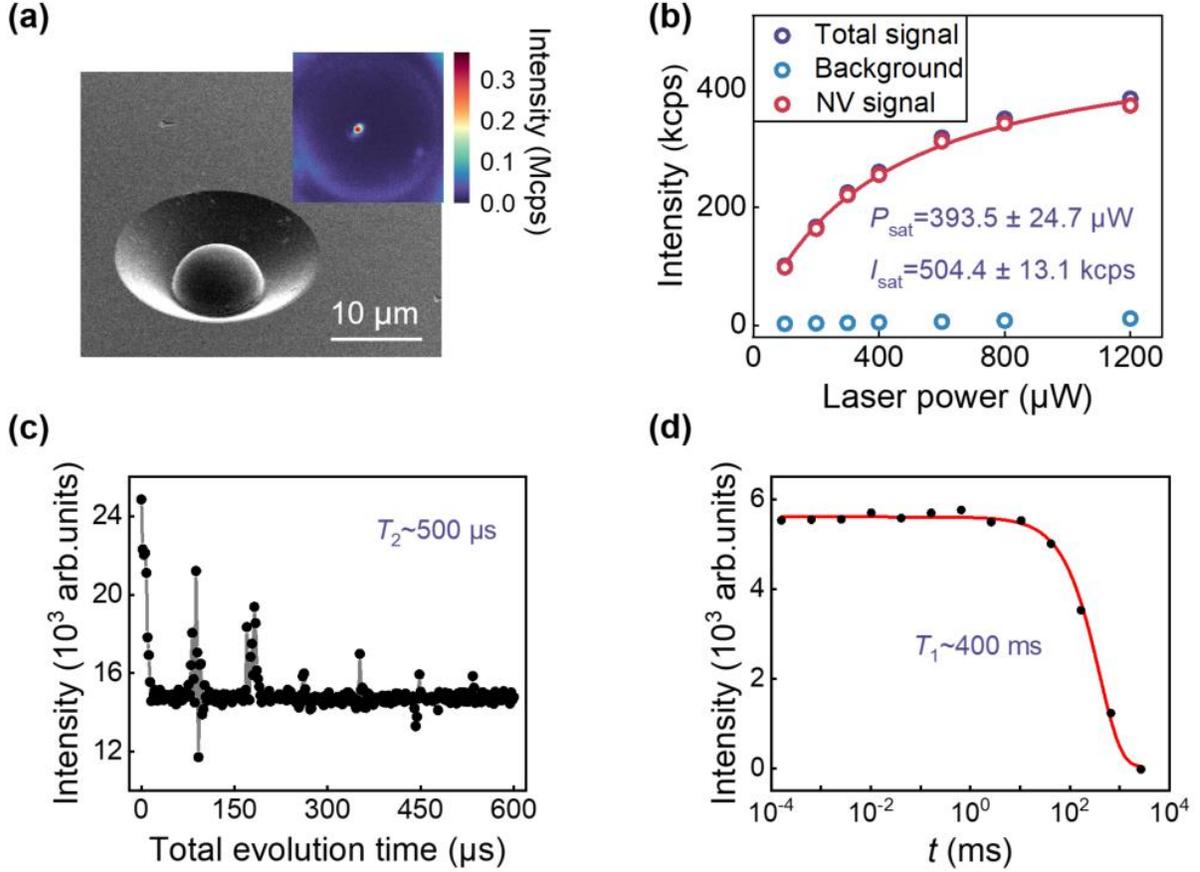

**FIG. 4. SIL fabrication and spin coherence of NV centers.** (a) SEM image of a solid immersion lens (SIL) etched on the diamond surface. Inset: confocal image of the NV center in the SIL. (b) Counting rate of the NV center as a function of the power of the 532-nm excitation laser. (c) Spin echo, and (d) spin relaxation signals of the measured NV center.

To improve the fluorescence collection efficiency of the NV center, we etched solid immersion lenses (SILs) above same measured NV centers [28,29]. Figure 4 (a) shows the scanning electron microscope (SEM) image of a SIL (radius 5 μm) on sample S1. The intensity of NV fluorescence is ~5-fold stronger than before etching (Fig. 4 (b)). We measured the PLE spectrum of this NV center in SIL. The splitting between $E_x$ and $E_y$ is 2.68 GHz, and their linewidths are 117.3 ± 0.9 MHz and 84.3 ± 2.0 MHz, respectively (see Fig. S2 for extended data). This result is comparable to that of unprocessed NV centers, indicating that the fabrication process does not cause additional lattice damage. We also etched SIL (radius 5 μm) on sample S2, but due to the relatively dense NV centers in this sample, the fluorescence of multiple NV centers is collected in the single SIL. See Fig. S4 in the Supplemental Material for more details.

Good spin coherence is also essential to advanced quantum applications. As shown in Fig. 4 (c-d), the NV centers produced by HPHT annealing has $T_2$~500 μs, and $T_1$~400 ms, which are consistent with the typical values of NV centers in high-purity diamonds. The collapse and revival behavior shown in Fig. 4 (e) indicates that the dominant mechanism of NV spin decoherence is the $^{13}$C nuclear spin bath [30]. The beating signal shown in Fig. 4 (c-d) is caused by a strongly coupled $^{13}$C with a coupling strength of 0.7 MHz. The NV centers in sample S2 exhibit similar spin coherence, as shown in Fig. S4. These experimental observations suggest that the HPHT annealing process does not introduce new impurity defects.

**Conclusions**

To summarize, we demonstrate that HPHT annealing of high-purity CVD diamonds enables the generation of NV centers with excellent optical and spin coherence. We applied a high pressure of 5.5 GPa to prevent graphitization of the diamond sample so that annealing can be performed at a high temperature and for hours, which greatly improves the flexibility in exploring the parameter space of NV center formation. The relatively uniform distribution of NV centers indicates that these color centers are formed by intrinsic N atoms and vacancies in the diamond lattice. The experimental results show that the generated NV centers exhibit excellent spin coherence. They also show a stable PLE spectrum with narrow linewidths (less than 100 MHz), comparable to the best reported results of natural NV centers in high-purity diamonds. Such good spin and optical coherence is essential for advanced quantum applications such as electron spin single-shot readout and entanglement of remote NV centers.

HPHT annealing provides a simple and sufficient method to generate diamond NV centers with excellent optical and spin properties. This method can be used to create high-quality color centers in SiC, hBN, and other wide-bandgap semiconductors. Compared to destructive methods such as ion implantation, laser writing and electron irradiation, HPHT annealing does not introduce additional defects into the diamond lattice. More importantly, the annealing process also partially removes the original defects (P1 centers, vacancies), resulting in a better local environment (less charge and less local strain). With this method, the NV density can be well controlled by controlling the annealing parameters. Along this direction, further experiments can be performed with higher annealing temperature or longer annealing time, as well as to study the dependence on the lattice orientation [19]. Another interesting experiment is to track the formation and diffusion of NV centers during repeated HPHT annealing, which could provide further insight into the diffusion dynamics of defects in diamond [31]. Meanwhile, the finalized NV density can give an indication of the initial nitrogen concentration and vacancies in diamond, which is still a challenge in the quantification of ultrapure diamond.

**Acknowledgments:** This project is supported by the Natural Science Foundation of Beijing, China (Grant Nos. Z200009, Z230005), the National Key Research and Development Program of China (Grants No. 2019YFA0308100), the National Natural Science Foundation of China (Grant Nos. 11974020, 12022509, 11934018, T2121001, 12374012), and the Chinese Academy of Sciences (Grant Nos. YJKYYQ20190082, XDB28030000).

## References

[1]  M. W. Doherty, N. B. Manson, P. Delaney, F. Jelezko, J. Wrachtrup, and L. C. L. Hollenberg, *The Nitrogen-Vacancy Colour Centre in Diamond*, Phys. Rep. **528**, 1 (2013).
[2]  L. Robledo, L. Childress, H. Bernien, B. Hensen, P. F. A. Alkemade, and R. Hanson, *High-Fidelity Projective Read-out of a Solid-State Spin Quantum Register*, Nature **477**, 574 (2011).
[3]  H. Bernien et al., *Heralded Entanglement between Solid-State Qubits Separated by Three Metres*, Nature **497**, 86 (2013).
[4]  B. Hensen et al., *Loophole-Free Bell Inequality Violation Using Electron Spins Separated by 1.3 Kilometres*, Nature **526**, 682 (2015).
[5]  M. Pompili et al., *Realization of a Multinode Quantum Network of Remote Solid-State Qubits*, Science **372**, 259 (2021).
[6]  R. Monge, T. Delord, G. Thiering, Á. Gali, and C. A. Meriles, *Resonant Versus Nonresonant Spin Readout of a Nitrogen-Vacancy Center in Diamond under Cryogenic Conditions*, Phys. Rev. Lett. **131**, 236901 (2023).
[7]  N. B. Manson, M. Hedges, M. S. J. Barson, R. Ahlefeldt, M. W. Doherty, H. Abe, T. Ohshima, and M. J. Sellars, *$NV^--N^+$ Pair Centre in 1b Diamond*, New J. Phys. **20**, 113037 (2018).
[8]  H. D. Robinson and B. B. Goldberg, *Light-Induced Spectral Diffusion in Single Self-Assembled Quantum Dots*, Phys. Rev. B **61**, R5086 (2000).
[9]  S. B. van Dam et al., *Optical Coherence of Diamond Nitrogen-Vacancy Centers Formed by Ion Implantation and Annealing*, Phys. Rev. B **99**, 161203 (2019).

[10] M. Kasperczyk, J. A. Zuber, A. Barfuss, J. Kölbl, V. Yurgens, S. Flågan, T. Jakubczyk, B. Shields, R. J. Warburton, and P. Maletinsky, *Statistically Modeling Optical Linewidths of Nitrogen Vacancy Centers in Microstructures*, Phys. Rev. B **102**, 075312 (2020).
[11] L. Orphal-Kobin, K. Unterguggenberger, T. Pregnolato, N. Kemf, M. Matalla, R.-S. Unger, I. Ostermay, G. Pieplow, and T. Schröder, *Optically Coherent Nitrogen-Vacancy Defect Centers in Diamond Nanostructures*, Phys. Rev. X **13**, 011042 (2023).
[12] E. Togan et al., *Quantum Entanglement between an Optical Photon and a Solid-State Spin Qubit*, Nature **466**, 730 (2010).
[13] P. C. Humphreys, N. Kalb, J. P. J. Morits, R. N. Schouten, R. F. L. Vermeulen, D. J. Twitchen, M. Markham, and R. Hanson, *Deterministic Delivery of Remote Entanglement on a Quantum Network*, Nature **558**, 268 (2018).
[14] Y.-C. Chen et al., *Laser Writing of Coherent Colour Centres in Diamond*, Nat. Photonics **11**, 77 (2017).
[15] M. Ruf, M. IJspeert, S. van Dam, N. de Jong, H. van den Berg, G. Evers, and R. Hanson, *Optically Coherent Nitrogen-Vacancy Centers in Micrometer-Thin Etched Diamond Membranes*, Nano Lett. **19**, 3987 (2019).
[16] K. C. Wong et al., *Microscopic Study of Optically Stable Coherent Color Centers in Diamond Generated by High-Temperature Annealing*, Phys. Rev. Appl. **18**, 024044 (2022).
[17] Y. Chu et al., *Coherent Optical Transitions in Implanted Nitrogen Vacancy Centers*, Nano Lett. **14**, 1982 (2014).
[18] J. O. Orwa et al., *Engineering of Nitrogen-Vacancy Color Centers in High Purity Diamond by Ion Implantation and Annealing*, J. Appl. Phys. **109**, 083530 (2011).
[19] X. Zhang et al., *Highly Coherent Nitrogen-Vacancy Centers in Diamond via Rational High-Pressure and High-Temperature Synthesis and Treatment*, Adv. Funct. Mater. 2309586 (2023).
[20] H. Lim, S. Park, H. Cheong, H.-M. Choi, and Y. C. Kim, *Discrimination between Natural and HPHT-Treated Type IIa Diamonds Using Photoluminescence Spectroscopy*, Diam. Relat. Mater. **19**, 1254 (2010).
[21] S. J. Charles, J. E. Butler, B. N. Feygelson, M. E. Newton, D. L. Carroll, J. W. Steeds, H. Darwish, C.-S. Yan, H. K. Mao, and R. J. Hemley, *Characterization of Nitrogen Doped Chemical Vapor Deposited Single Crystal Diamond before and after High Pressure, High Temperature Annealing*, Phys. Status Solidi (a) **201**, 2473 (2004).
[22] L. Dei Cas, S. Zeldin, N. Nunn, M. Torelli, A. I. Shames, A. M. Zaitsev, and O. Shenderova, *From Fancy Blue to Red: Controlled Production of a Vibrant Color Spectrum of Fluorescent Diamond Particles*, Adv. Funct. Mater. **29**, 1808362 (2019).
[23] J. Meijer, B. Burchard, M. Domhan, C. Wittmann, T. Gaebel, I. Popa, F. Jelezko, and J. Wrachtrup, *Generation of Single Color Centers by Focused Nitrogen Implantation*, Appl. Phys. Lett. **87**, 261909 (2005).
[24] S. Pezzagna and J. Meijer, *Quantum Computer Based on Color Centers in Diamond*, Appl. Phys. Rev. **8**, 011308 (2021).
[25] S. Onoda, K. Tatsumi, M. Haruyama, T. Teraji, J. Isoya, W. Kada, T. Ohshima, and O. Hanaizumi, *Diffusion of Vacancies Created by High-Energy Heavy Ion Strike Into Diamond*, Phys. Status Solidi (a) **214**, 1700160 (2017).
[26] A. Batalov, V. Jacques, F. Kaiser, P. Siyushev, P. Neumann, L. J. Rogers, R. L. McMurtrie, N. B. Manson, F. Jelezko, and J. Wrachtrup, *Low Temperature Studies of the Excited-State Structure of Negatively Charged Nitrogen-Vacancy Color Centers in Diamond*, Phys. Rev. Lett. **102**, 195506 (2009).
[27] J. R. Maze, A. Gali, E. Togan, Y. Chu, A. Trifonov, E. Kaxiras, and M. D. Lukin, *Properties of Nitrogen-Vacancy Centers in Diamond: The Group Theoretic Approach*, New J. Phys. **13**, 025025 (2011).
[28] J. P. Hadden, J. P. Harrison, A. C. Stanley-Clarke, L. Marseglia, Y.-L. D. Ho, B. R. Patton, J. L. O'Brien, and J. G. Rarity, *Strongly Enhanced Photon Collection from Diamond Defect Centers under Microfabricated Integrated Solid Immersion Lenses*, Appl. Phys. Lett. **97**, 241901 (2010).
[29] M. Jamali, I. Gerhardt, M. Rezai, K. Frenner, H. Fedder, and J. Wrachtrup, *Microscopic Diamond Solid-Immersion-Lenses Fabricated around Single Defect Centers by Focused Ion Beam Milling*, Rev. Sci. Instrum. **85**, 123703 (2014).
[30] L. Childress, M. V. Gurudev Dutt, J. M. Taylor, A. S. Zibrov, F. Jelezko, J. Wrachtrup, P. R. Hemmer, and M. D. Lukin, *Coherent Dynamics of Coupled Electron and Nuclear Spin Qubits in Diamond*, Science **314**, 281 (2006).
[31] M. N. R. Ashfold, J. P. Goss, B. L. Green, P. W. May, M. E. Newton, and C. V. Peaker, *Nitrogen in Diamond*, Chem. Rev. **120**, 5745 (2020).